\documentclass[12pt,aps,prb,preprint,showpacs]{revtex4-1}
\usepackage{bm,color,amsmath,amssymb,mathrsfs,latexsym,graphicx,psfrag}

\def\be{\begin{equation}}
\def\ee{\end{equation}}
\def\ba{\begin{eqnarray}}
\def\ea{\end{eqnarray}}

\begin{document}

\title{Spin-orbit coupling and odd-parity superconductivity in the
quasi-one-dimensional compound
Li$_{0.9}$Mo$_6$O$_{17}$}
\author{Christian Platt${}^1$, Weejee Cho${}^1$, Ross H. McKenzie${}^2$, Ronny Thomale${}^{1,3}$, and S. Raghu$^{1,4}$}

\affiliation{${}^1$ Department of Physics, Stanford University, Stanford, CA 94305, USA\\
${}^2$ School of Mathematics and Physics, University of Queensland, Brisbane, 4072 Queensland, Australia\\
${}^3$ Institute for Theoretical Physics, University of Wuerzburg, D-97074 Wuerzburg, Germany\\
${}^4$ SLAC National Accelerator Laboratory, Menlo Park, California 94025, USA
}

\date{\today}
\begin{abstract}
Previous theoretical studies [W. Cho, C. Platt, R. H. McKenzie, and S. Raghu, Phys. Rev. B 92, 134514 (2015); N. Lera and J. V. Alvarez, Phys. Rev. B 92, 174523 (2015)] have suggested that Li$_{0.9}$Mo$_6$O$_{17}$, a quasi-one dimensional  ``purple bronze" compound, exhibits spin-triplet superconductivity and that the gap function changes sign across the two nearly degenerate Fermi surface sheets. 
We investigate the role of spin-orbit coupling (SOC) in determining the symmetry
and orientation of the $d$-vector associated with the
superconducting order parameter.
We propose that the lack of local inversion symmetry within
the four-atom unit cell leads to a staggered spin-orbit
coupling analogous to that proposed
for graphene, MoS$_2$, or SrPtAs.
In addition, from a weak-coupling renormalization group treatment of an effective model Hamiltonian, we find that SOC favors the odd parity
$A_{1u}$ state with $S_z = \pm 1$ over the $B$ states with $S_z=0$, where $z$ denotes the least-conducting direction.  We discuss possible definitive experimental signatures of this superconducting state.
\end{abstract}

\pacs{71.10.Fd, 71.10.Hf, 71.27.+a, 74.20.Rp}
\maketitle

\emph{Introduction.---}
In conventional superconductors, the spin-degree of freedom is frozen due to the singlet nature of  Cooper pairs.
However, in certain unconventional superconductors, the spin-degree of freedom remains active when pairing involves
the formation of triplet states.  The most familiar example is superfluid He$^3$, in which several spin-triplet states occur.
The order parameter has a richer structure in such systems, which in turn leads to more subtle collective modes and
topological excitations.  Consequently, many fascinating experimental signatures (e.g., in NMR) of triplet superconductivity
have been proposed and identified 
 in a diverse range of materials including K$_2 $Cr$_3$As$_3$ \cite{Bao2015,Zhi2015},
TMTSF$_2$X \cite{lebed2000,Zhang2007}, strontium ruthenate \cite{Maeno2012},
and the heavy fermion compound UPt$_3$ \cite{strand2010}.

In a spin-triplet superconductor, spin-orbit coupling (SOC) can have a qualitative effect on the nature of the ground state.
This is true even in a neutral superfluid such as He$^3$, where spin-orbit effects due to dipole-dipole forces can lock the relative orientation of spin and orbital angular momentum 
of the order parameter~\cite{Legget75}.  It follows that spin-orbit effects can play an even more  vital role in
many correlated electron materials
that exhibit spin-triplet superconductivity. 
As $SU(2)$ spin symmetry
is broken due to spin-orbit effects, generically one cannot speak of a ``spin-triplet'' state; instead, if the material retains inversion
symmetry (parity) in the normal state---as is the case in the present study---one may refer to odd-parity superconductivity, in
which the Cooper pair wave-function is odd under inversion.

There are different perspectives on studying the effects of SOC on odd-parity superconductivity. As a more
phenomenological approach, one takes symmetry considerations into account and studies the role of spin-orbit effects near the superconducting transition.  Such considerations, based on Landau-Ginzburg theory, inform us on the possible nature
of the ground states by enumerating the set of irreducible representations consistent with the symmetries of the normal state~\cite{Sigrist91,volovik1993,Mineev99}.  Only a more microscopic theory, 
which takes
into account the interplay between SOC and interactions, can predict which of these allowed state is the favored
ground state. The microscopic approach to unconventional superconductivity, taking into account both electron interactions
and spin-orbit physics, has been a persisting challenge~\cite{scaffidi2014,wang2013,yanase2014}.  Here, we explore such effects in the context of Li$_{0.9}$Mo$_6$O$_{17}$,
a layered, quasi-one-dimensional material known more commonly as a ``purple bronze."

\begin{figure}[t]
\includegraphics[scale=0.18]{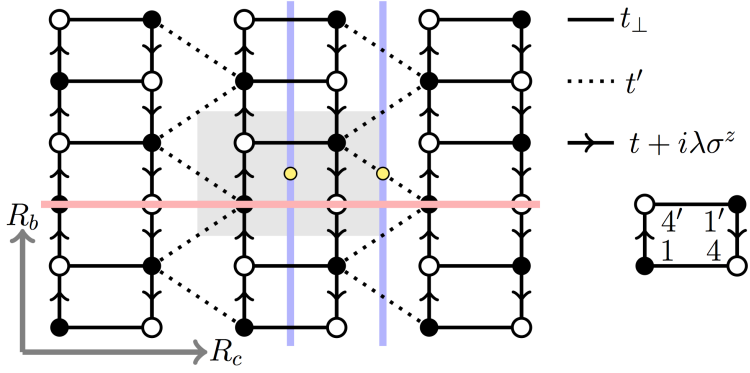}\\
\caption{Tight-binding lattice model. Each circle corresponds to a single Mo atom and there are four atoms per unit cell (gray rectangle).
Filled and empty circles denote two types of crystallographically inequivalent Mo atoms, i.e., Mo(1) and Mo(4), within the layers of Li$_{0.9}$Mo$_6$O$_{17}$.
Intra-chain, intra-ladder and inter-ladder hopping integrals are denoted $t$, $t_\perp$, and $t'$, respectively. SOC is represented by a
spin-dependent hopping term $+i\lambda\sigma^z$ ($-i\lambda\sigma^z$) along (opposite)
to the arrow directions within the chains. The two yellow circles define centers of $C_2$ rotational
symmetry, the blue (red) lines indicate glide mirror (mirror) planes. 
}
\label{ladder}
\end{figure}

There are several indications that this material likely exhibits spin-triplet pairing, among them the display of a pronounced anisotropy of the
upper critical field.  In particular, the upper critical field along the crystallographic $b$ axis 
exceeds the 
Chandrasekhar-Clogston limit, which both suggests
the possibility of spin-triplet pairing and highlights the important role of SOC~\cite{mercure2012}.  Motivated by these
and other experiments~\cite{dudy2013,podlich,matsuda1986} that point towards unconventional superconductivity, we have studied a weak coupling limit of a model Hamiltonian suggested for this system in a previous paper~\cite{mckenzie2012prb}.
Our results indicated that a triplet state with accidental nodes was indeed favored over singlet states~\cite{Cho2015} (see also Ref.~\onlinecite{Lera2015}).
Here, we refine our analysis 
to investigate how the spin degeneracy of the triplet state is lifted in the presence of SOC. We construct a
SOC Hamiltonian that is consistent with the symmetries of the model and study the superconducting instabilities as a
function of the SOC coupling constant. Our main results can be summarized as follows: defining the $z$ direction to be perpendicular to the plane
(the least-conducting direction) in Fig.~\ref{ladder}, we find that SOC favors
an $S_z=\pm 1$ triplet pairing state, corresponding to an in-plane $d$-vector orientation. 
This result is independent of the sign of the SOC constant, as we show below.

\emph{Electronic structure considerations. ---}
The low-energy electronic degrees of freedom in Li$_{0.9}$Mo$_6$O$_{17}$ reside on two-leg ladders built from the $d_{xy}$ orbitals of Mo atoms.  Here, the constituting chains
run along the crystallographic $b$ axis and are weakly coupled along the $c$ direction via $t_{\perp}$ and $t^{\prime}$ as shown in Fig.~\ref{ladder}. In units of the intra-chain hopping amplitude $t$, we set $t_{\perp}=-0.048\eta_{w}t$ and $t^{\prime}=0.072\eta_{w}t$, where $\eta_{w}$ is an additional parameter controlling the Fermi-surface warping and nesting properties.
As the results of our calculations only differ in minor details for $\eta_{w}$ in a range of $0.5<\eta_{w}<1.5$, we set $\eta_{w} = 1.0$ in accordance with Ref.~\onlinecite{mckenzie2012prb}.
Similar, but slightly different tight-binding models have been presented in Refs.~\onlinecite{chudzinski2012} and~\onlinecite{nuss2014}.

\emph{Tight-binding model.---}
As a minimal effective Hamiltonian of the low-energy electronic properties, we consider a Hubbard model \cite{mckenzie2012prb} near
quarter filling ($n_{\textmd{el}}=1.9$ out of 8 per unit cell).
The tight-binding part of the Hamiltonian $\mathcal{H}$ is
\begin{equation}\label{eq:ham1}
\mathcal{H}_0 = \sum_k C^{\dagger}_{k}H(k)C^{\phantom{\dagger}}_{k}
\end{equation}
with $C_{k} = ( c_{k1s}, c_{k4s}, c_{k1's}, c_{k4's} )^T$. We divide the tight-binding Hamiltonian into the kinetic term $H_0(k)$ and the SOC term $H_{\text{soc}}(k)$ such that $H(k) = H_0(k) + H_{\text{soc}}(k)$. The kinetic term reads
\begin{equation}
H_0(k) =
- \begin{pmatrix}
0                                           & t_{\perp}                                                        & t'e^{-ik_x}(1 + e^{-ik_y})                                   & t(1 + e^{-ik_y})\\
t_{\perp}                               &  0                                                                   & t(1 + e^{-ik_y})                                                  & 0 \\
t'e^{ik_x}(1 + e^{ik_y})         &  t(1 + e^{k_y})                                                & 0                                                                       &  t_{\perp} \\
t(1 + e^{ik_y})                       &  0                                                                   & t_{\perp}                                                           &  0 \\
\end{pmatrix}
\otimes \sigma_0.
\end{equation}
Here, the $x$, $y$, and $z$ directions correspond to the crystalline $c$, $b$, and $-a$ directions, respectively.

\emph{Spin-orbit coupling.—}
We include spin-orbit interactions in the form of a nearest-neighbor spin-dependent
hopping amplitude $\pm i\lambda\sigma^z$ along the chains. Here, the different signs correspond to hopping directions
along and opposite to the bond arrows depicted in Fig.~\ref{ladder}.  Within our model description, this type of spin-orbit interaction
originates from the lack of reflection symmetry across a single chain. More precisely, this lack of reflection symmetry gives rise to a net electric field
perpendicular to the chains, which in turn couples the electron's propagation to its spin.
Since the low energy dynamics arises from a single orbital ({\it i.e.} the $d_{xy}$ orbital), atomic angular momentum is quenched in this system, and the atomic spin-orbit coupling of the form $H_{\text{a-SOC}} \sim \vec L \cdot \vec S$ does not arise.  Consequently, the allowed form of SOC must depend on the Bloch wavevector.  To determine the symmetry-allowed coupling, we note from Fig.~\ref{ladder} that our model possesses a horizontal mirror plane and a vertical glide mirror plane. 
Requiring that 1) {\it both} these planes of symmetry 
be preserved and 2) that the normal state retains inversion symmetry, and 3) recalling that the spin is an axial vector, we are led to two conclusions.  First, Rashba SOC, which requires bulk inversion symmetry breaking, cannot occur: the pattern of spin-orbit coupling must be staggered.  Secondly, such SOC can only involve the $z$ component of the spin: if either the $x, y$ components were involved, the planes of symmetry 
described above would be lost.  Thus, symmetry considerations constrain SOC to be of the form
\begin{equation*}
H_{\text{soc}}(k)=
\begin{pmatrix}
0                                  &                                   0 &                                   0 & -i\lambda(1 + e^{-ik_y}) \\
0                                  &                                   0 & i\lambda(1 + e^{-ik_y}) &                                     0 \\
0                                  & -i\lambda(1 + e^{ik_y}) &                                   0 &                                     0 \\
i\lambda(1 + e^{ik_y}) &                                   0 &                                   0 &                                     0 \\
\end{pmatrix}
\otimes\sigma_z.
\end{equation*}
The $(4\times 4)$-matrices in the above notation act in the space of four inequivalent Mo atoms in the unit cell (Fig.~\ref{ladder}), whereas the Pauli-matrices $\sigma_0$, $\sigma_z$ only affect the spin degree of freedom. The only symmetries explicitly broken by $H_{\text{soc}}$ are the spin-rotational symmetries generated by $\sigma_x$ and $\sigma_y$. All other symmetries, such as inversion, time-reversal, and spin-rotation symmetry around $z$, are still intact and will be used to classify the different pairing states. Such a form of SOC  is reminiscent of that present in materials with ``local inversion symmetry breaking,'' as described in Refs.~\onlinecite{Fischer2011,maruyama2012}.  The basic idea is 
that while the material does possess inversion symmetry, one or more sites per unit cell do not coincide with inversion centers. 
Other examples include graphene, where each sublattice locally breaks inversion but the triangular Bravais lattice is manifestly centrosymmetric
\cite{Kane2005,Min2006}, monolayers of
MoS$_2$ \cite{Xiao2012}, and the new pnictide superconductor SrPtAs \cite{Youn2012}, where the latter has recently been suggested to host chiral singlet superconductivity~\cite{PhysRevB.89.020509}.

We assume an on-site
repulsion term
\begin{equation}\label{eq:onsite}
\mathcal{H}_{int} = U \sum_{i}\sum_{o}n_{io\uparrow}n_{io\downarrow}
 = \frac{U}{N}\sum_{\{k_{i}\}}\sum_{o}c^{\dagger}_{k_1o\uparrow}c^{\dagger}_{k_2o\downarrow}c^{\phantom{\dagger}}_{k_4o\downarrow}c^{\phantom{\dagger}}_{k_3o\uparrow},
\end{equation}
where momentum conservation $k_4 = k_1 + k_2 - k_3$ is implicitly imposed, $N$ denotes the number of unit-cells, and $o$ labels the four inequivalent Mo sites. The model
we consider reads
\begin{equation}
\mathcal{H} = \mathcal{H}_0 + \mathcal{H}_{int}
\end{equation}
and is most conveniently, at least in the case of weak coupling with $|t|$, $|t_{\perp}|$, $|t'|$, $|\lambda| \gg U$, represented in an eigenbasis of $\mathcal{H}_0$:
\begin{equation}
\gamma^{\dagger}_{kb s} = \sum_{o} a^s_{b o}(k)c^{\dagger}_{kos}.
\end{equation}
Here, the corresponding states $|kbs\rangle = \gamma^{\dagger}_{kb s}|0\rangle$ fulfill $\mathcal{H}_0|kbs\rangle = \epsilon_b(k)|kbs\rangle$ and can still be labeled by the $S_z$ quantum number $s$.
The band index $b=1,\ldots,4$ enumerates the corresponding energy bands $\epsilon_b(k)$, which are at least two-fold degenerate due to combined inversion and time-reversal symmetry.
The effect of the spin-orbit coupling $\lambda$ is small on the band structure $\epsilon_b(k)$ 
but rather significant on the states $|kbs\rangle$. The resulting model in the band basis then reads
\begin{equation}\label{bandmodel}
\mathcal{H} =
\sum_{kbs} \epsilon_b(k)\gamma^{\dagger}_{kbs}\gamma^{\phantom{\dagger}}_{kbs} +
\sum_{\{k_i,b_i\}} V(k_1b_1, k_2b_2, k_3b_3, k_4b_4)
\gamma^{\dagger}_{k_1b_1s_1}\gamma^{\dagger}_{k_2b_2s_2}\gamma^{\phantom{\dagger}}_{k_4b_4s_4}\gamma^{\phantom{\dagger}}_{k_3b_3s_3},
\end{equation}
where $k_4 = k_1 + k_2 - k_3$ modulo reciprocal lattice vectors, and  the coupling function given by
\begin{equation}
V(k_1b_1, k_2b_2, k_3b_3, k_4b_4) =  \frac{U}{N}\sum_{o}a^{\uparrow}_{b_1o}(k_1)a^{\downarrow}_{b_2o}(k_2)a^{\uparrow*}_{b_3o}(k_3)a^{\downarrow*}_{b_4o}(k_4).
\end{equation}

\emph{Constraints from symmetry.---}
Before proceeding with the weak-coupling solution, we wish to outline the possible superconducting states that may arise based on symmetry considerations alone.
In addition to possessing time-reversal and spatial inversion symmetry, the Hamiltonian is invariant under 
1) a $U(1)$ spin rotation about the $z$ axis and 2)
reflections about the $xy$. $yz$, and $zx$ planes, and 3) $\pi$ rotations about the $x$, $y$, and $z$ axes.
Although in a strict sense the real-space lattice model only has glide reflection symmetry about the $yz$ plane, 
the $k$-space Hamiltonian in Eq. (\ref{bandmodel}) is symmetric under the reflection about this plane 
due to an appropriate basis choice that incorporates
additional Bloch phases. The point-group of $\mathcal{H}$ in Eq. (\ref{bandmodel}) is therefore $D_{2h}$. Note that while the $U(1)$ rotation transforms the spin alone, the reflections transform both the spin and momentum components. Specifically, the reflection about the $yz$ plane acts on a $\textbf{k}$-dependent spin-1/2 object as
\begin{equation}
\label{reflectionop}
\tau_{x}: f(k_x, k_y, k_z)\,|s\rangle \to \pm i f(-k_x, k_y, k_z)\,\sigma_{x}|s\rangle
\end{equation}
($|s\rangle$ denotes a spin state), and similarly for other reflections $\tau_{y}$ and $\tau_{z}$.

Having enumerated the symmetries of the normal state, and neglecting $k_z$ dependence of the order parameter, it follows that there are four 
distinct irreducible representations corresponding to odd-parity superconductivity in our model~\cite{annett1990,lebed2000}:
\begin{eqnarray}
&&A_{1u}: \ \ \ \bm d(\bm k) = \eta_x(\bm k) \hat x + \alpha \eta_y(\bm k) \hat y, \nonumber \\
&&B_{1u}: \ \ \ \bm d(\bm k) = \alpha\eta_y(\bm k) \hat x + \eta_x(\bm k) \hat y, \nonumber \\
&&B_{2u}: \ \ \ \bm d(\bm k) = \eta_x(\bm k) \hat z, \nonumber \\
&&B_{3u}: \ \ \ \bm d(\bm k) =  \eta_y(\bm k) \hat z.
\end{eqnarray}
Here, $\hat{x}$, $\hat{y}$, and $\hat{z}$ respectively denote the triplet states proportional to $-\vert\uparrow\uparrow\rangle + \vert\downarrow\downarrow\rangle$, $\vert\uparrow\uparrow\rangle + \vert\downarrow\downarrow\rangle$, and $\vert\uparrow\downarrow\rangle + \vert\downarrow\uparrow\rangle$; $\vec \eta(\bm k)$ is a function of momentum that transform as the components of momentum (e.g., $\eta_i = \sin{\left( k_i \right)}$ with the lattice constants set to unity); 
$\alpha$ is an arbitrary real-valued constant. In our symmetry analysis, we neglect an overall complex factor that is always present in the superconducting order parameter which has no observable consequences.

The $A_{1u}$ and $B_{1u}$ representations have in-plane $d$-vectors, which correspond to linear combinations of states with $S_z = \pm 1$. Rotation of the spin about the $z$ axis, which is a symmetry operation, mixes these representations, thereby rendering them degenerate. This degeneracy would be lifted if the normal state did not conserve $S_{z}$.  Under the reflection $\tau_z$ [see Eq. (\ref{reflectionop})], both $A_{1u}$ and $B_{1u}$ change sign; on the other hand, the two representations are respectively odd and even under $\tau_y$.  
The $B_{2u}$ and $B_{3u}$  representations
have the $d$-vector along the $z$ axis.  They can be thought of as triplet states with $S_z = 0$. They are both invariant under the reflection $\tau_{z}$. Under $\tau_{y}$, $B_{2u}$ is odd, whereas $B_{3u}$ is even. 


As we shall demonstrate below, our microscopic theory leads to the conclusion that the states with the $d$-vector oriented in-plane is favored.  In the presence of $S_z$ conservation, this
order parameter would have soft collective fluctuations corresponding to the freedom to ``rotate'' into an arbitrary linear superposition of the $A_{1u}$, $B_{1u}$ representations.
If $S_z$ conservation were broken - due to effects that are not captured in our present model---these representations would split, and the associated collective modes would be gapped.

\emph{Perturbative renormalization group (RG). ---}
Starting from the model Hamiltonian in (\ref{bandmodel}), we implemented an RG method~\cite{raghu2010weakcoupling,Raghu2010,PhysRevB.86.121105,cho2013} to investigate superconducting instabilities {\`a} la
Kohn and Luttinger \cite{kohn-65prl524}. The idea is to assume sufficiently small interactions such that a renormalized interaction near the Fermi surface can be safely
calculated by perturbation theory and still remains in a weak-coupling range. For the remainder modes, a standard RG procedure~\cite{Polchinski99,Shankar94}
is applied and gives significant renormalization only for couplings in the Cooper channel. This of course only holds if the system is not at a highly fine tuned point of the phase diagram
at which even infinitesimally small interactions induce other competing channels.
\begin{figure}[t]
\includegraphics[scale=0.25]{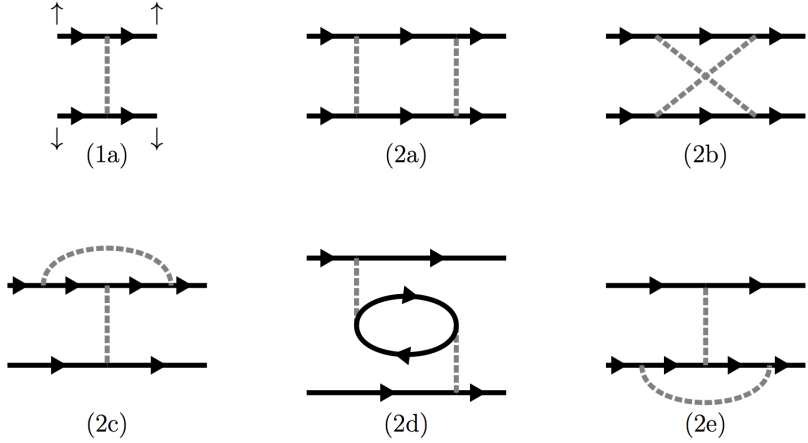}\\
\caption{First and second order contribution to the effective interaction $V_{\text{eff}}$. Each solid line corresponds to a propagator with
momentum $k$, band-index $b$ and spin $s$. The dashed line refers to the interaction in~(\ref{bandmodel}) and provides nonzero contributions
for spin configurations of the type indicated in diagram (1a).}
\label{diag_fig}
\end{figure}
As a first step, we therefore determine the effective interaction $V_{\text{eff}}$ at energy scales close to the Fermi surface by calculating the lowest order diagrams shown in Fig.~\ref{diag_fig}.
Before we proceed with the subsequent RG treatment in the Cooper channel, it is useful to organize the pair scattering in terms of irreducible representations of $S_z$ and parity
\begin{align}\nonumber
\mathcal{H}_{\text{int}}^{\text{pair}}&=\frac{1}{2}\sum_{k,q}\sum_{\{s_i\}}
V_{\text{eff}}(ks_1,-ks_2,qs_3,-qs_4)\gamma^{\dagger}_{ks_1}\gamma^{\dagger}_{-ks_2}
\gamma^{\phantom{\dagger}}_{-qs_4}\gamma^{\phantom{\dagger}}_{qs_3}\\\label{decomp}
&= \frac{1}{2}\sum_{k,q}\left[\Gamma_{0}(k,q)\psi^{\dagger}_{0,k}\psi^{\phantom{\dagger}}_{0,q} + \Gamma_{1}(k,q)\psi^{\dagger}_{1,k}\psi^{\phantom{\dagger}}_{1,q}
+ \Gamma_{2}(k,q)\psi^{\dagger}_{2,k}\psi^{\phantom{\dagger}}_{2,q} + \Gamma_{3}(k,q)\psi^{\dagger}_{3,k}\psi^{\phantom{\dagger}}_{3,q}\right].
\end{align}
Here, we used the following notation for the pairing operators
\begin{align}\nonumber
\psi^{\dagger}_{0,k} &= \frac{1}{\sqrt{2}}\left(\gamma^{\dagger}_{k\uparrow}\gamma^{\dagger}_{-k\downarrow} - \gamma^{\dagger}_{k\downarrow}\gamma^{\dagger}_{-k\uparrow}\right),
\quad\psi^{\dagger}_{1,k} = \gamma^{\dagger}_{k\uparrow}\gamma^{\dagger}_{-k\uparrow},\\\nonumber
\psi^{\dagger}_{2,k} &= \frac{1}{\sqrt{2}}\left(\gamma^{\dagger}_{k\uparrow}\gamma^{\dagger}_{-k\downarrow} + \gamma^{\dagger}_{k\downarrow}\gamma^{\dagger}_{-k\uparrow}\right),
\quad\psi^{\dagger}_{3,k} = \gamma^{\dagger}_{k\downarrow}\gamma^{\dagger}_{-k\downarrow},
\end{align}
and omitted the band indices $b$ which are implicitly included in the momentum $k$ because all modes away from the Fermi surface have been scaled out.
The various coupling functions $\Gamma$ in (\ref{decomp}) can be inferred from the effective interaction $V_{\text{eff}}$ or, respectively, the diagrams Fig.~\ref{diag_fig}:
\begin{align}\nonumber
\Gamma_0(k,q) &= \frac{1}{2}\Big[d^{1a}(k\uparrow,-k\downarrow,q\uparrow,-q\downarrow)+ d^{2a}(k\uparrow,-k\downarrow,q\uparrow,-q\downarrow)\\\nonumber
 &\quad+ d^{2b}(k\uparrow,-k\downarrow,q\uparrow,-q\downarrow) +  (k \leftrightarrow -k)\Big],\\\nonumber
\Gamma_1(k,q) & = \frac{1}{2}\Big[d^{2d}(k\uparrow,-k\uparrow,q\uparrow,-q\uparrow) -  (k \leftrightarrow -k)\Big]\\\nonumber
\Gamma_2(k,q) &= \frac{1}{2}\Big[d^{2b}(k\uparrow,-k\downarrow,q\uparrow,-q\downarrow) -  (k \leftrightarrow -k)\Big]\\\nonumber
\Gamma_3(k,q) & = \frac{1}{2}\Big[d^{2d}(k\downarrow,-k\downarrow,q\downarrow,-q\downarrow) -  (k \leftrightarrow -k)\Big]
\end{align}
Here, the on-site nature of the bare interaction leads to a number of consequences: first, the diagrams (2c) and (2e) identically vanish; second, (2a) is also an on-site interaction; third, as both (1a) and (2a) are even in $k$, they do not contribute to $\Gamma_{1,2,3}$, which are odd in $k$.
If we further decompose the different coupling functions $\Gamma_i$ into eigenmodes defined by the integral equation
along the Fermi-surface
\begin{equation}\label{eigmode}
\oint \frac{d\hat{k}}{(2\pi)v_F(\hat{k})} \Gamma_i(\hat{k},\hat{q}) g_{ni}(\hat{q})= \lambda_{ni} g_{ni}(\hat{k}),
\end{equation}
the 1-loop RG flow in the Cooper channel decouples into separate flow equations for each~$\lambda_{ni}$
\begin{equation}\label{flow}
\frac{d\lambda_{ni}(l)}{dl} = -\lambda_{ni}^2(l),\quad \lambda_{ni}(l) = \frac{\lambda_{ni}(0)}{1 + \lambda_{ni}(0)l}.
\end{equation}
The initial values $\lambda_{ni}(0)$ are given by the eigenvalues in (\ref{eigmode}) and the index $ni$ labels the $n$-th eigenmode of $\Gamma_i$.
It is easy to see from (\ref{flow}) that a negative eigenvalue grows further under renormalization and that the most negative one $\lambda_{0i}$ eventually causes a pairing instability
with a transition temperature $T_c \sim W e^{-1/|\lambda_{0i}|}$ and a superconducting gap structure determined by the corresponding eigenmode $g_{0i}(\hat{k})$. It should also be noted that,
for asymptotically small interactions, the bare coupling of (1a) in Fig.~\ref{diag_fig} provides an infinitely larger contribution than the other terms (2a-e) and that (2a) has precisely the same momentum dependence as (1a). Then, for the purpose of calculating negative eigenvalues, one can simply project $\Gamma^{0}$ onto the null space of (1a) [and hence of (2a)] as discussed in more detail in Ref.~\onlinecite{Cho2015}. 

\begin{figure}
\centering
\includegraphics[scale = 0.6]{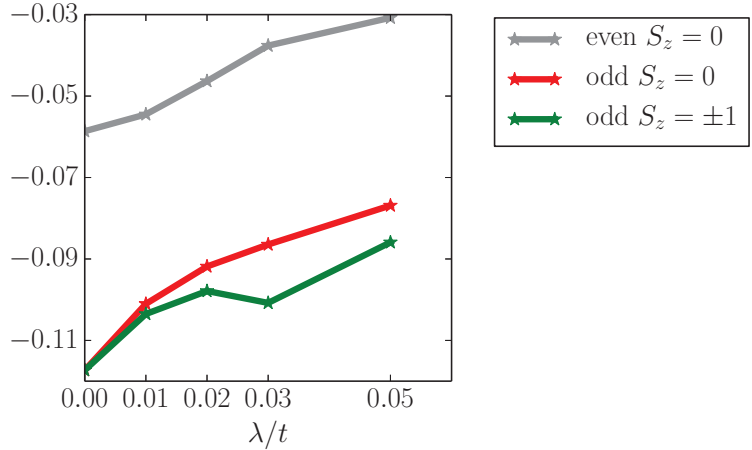}
\caption{Leading eigenvalues for the even-parity, odd-parity
$A_{1u}$ ($S_z=0$) and $B_{2u}$
($S_z=\pm 1$) channels as a function of the spin-orbit coupling $\lambda$.}
\label{pairing}
\end{figure}

\emph{Results of the weak-coupling analysis.---}
Figure~\ref{pairing} shows the dominant pairing strength in the different pairing channels as a functions of spin-orbit coupling $\lambda$.
As in Ref.~\onlinecite{Cho2015}, we have plotted the dimensionless quantity $\tilde{\lambda}_{0i} \equiv \lambda_{0i} \frac{W^{2}}{U^{2}}$ and only show data
for the band structure parameters corresponding to $\eta_w = 1$. Note that $\lambda$ without subindices denotes spin-orbit coupling and $W$ is the electronic bandwidth.

\begin{figure}
\centering
\includegraphics[scale = 0.55]{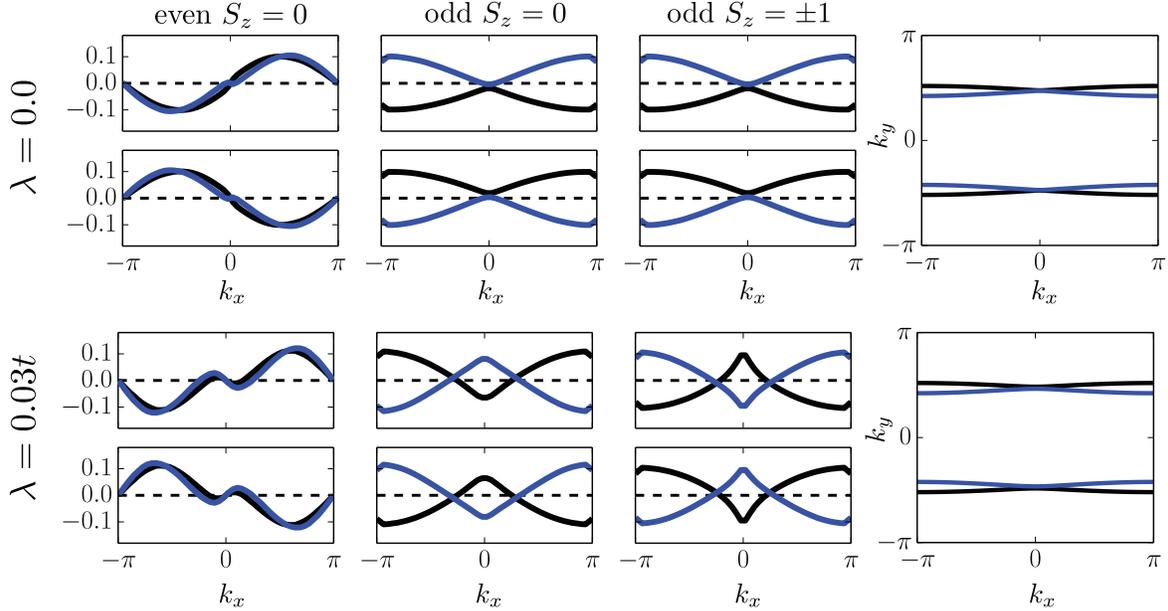}
\caption{(left to right) Leading eigenvectors in the even parity ($S_z=0$),  odd parity $A_{1u}$ ($S_z=0$) and $B_{2u}$ ($S_z=\pm 1$) channel. The upper row displays the case of zero spin-orbit
coupling $\lambda = 0.0t$, the lower one shows $\lambda = 0.03t$. The upper (lower) subdivision in each viewgraph corresponds to the upper (lower) Fermi surface of equal color coding.
}
\label{soc}
\end{figure}

Using an independent numerical implementation, we reproduced the results of our previous work \cite{Cho2015} in the limit of vanishing SOC $\lambda$.
Here, the odd-parity channel is clearly favored as compared to the even-parity one.  This conclusion also persists in the regime of finite SOC,
where the odd-parity state with total $S_z = \pm 1$ is preferred over the one with $S_z = 0$. Note that states with $S_z = 1$ and $S_z = -1$
are degenerate in terms of their eigenvalues in Eq.~(\ref{eigmode}) due to time-reversal symmetry and that in the limit of $\lambda\rightarrow 0$, also the
$S_z = 0$ channel merges as required by spin-rotation symmetry. As a general trend, it appears that the absolute eigenvalues in Fig.~\ref{pairing}, and with that also $T_c$,
decreases with increasing SOC. The associated pair wave functions $g_{0i}(\hat{k})$ along the Fermi surface are shown in Fig.~\ref{soc} for $\lambda = 0.0t$
and $\lambda = 0.03t$. Notice that when SOC is absent, the odd-parity solution exhibits a gap minimum at $k_x=0$ on each Fermi surface sheet (a more careful inspection reveals that the ``gap minimum'' in each of the inner fermi surfaces is a pair of closely spaced nodes); with increasing SOC, each gap minimum turns into a pair of nodes.  

\emph{Magnitude of the spin-orbit coupling $\lambda$. ---}
On a microscopic level the term $H_{\text{soc}}$ results from a perturbative treatment of the full atomic spin-orbit orbit interaction
$H_{\text{a-SOC}} \sim \vec{L}\cdot\vec{S}$ within the subspace of $d_{xy}$ states.
Obtaining a reliable estimate of $\lambda$ is subtle and
requires {\it ab initio} calculations, as shown
by Min et al.~\cite{Min2006} for graphene and
Xiao et al.~\cite{Xiao2012} for
MX$_2$ (M = Mo, W; X = S, Se).
Such microscopic calculations are beyond the scope of this study.
Instead, we varied $\lambda$ in a broad range and postpone a  microscopic calculation of
and estimate of $\lambda$ for this compound to a future study.

\emph{Discussion. ---}
In this paper, we have incorporated the effects of spin-orbit coupling in a weak-coupling treatment of superconductivity in purple bronze.  We have constructed a spin-orbit Hamiltonian by requiring that the reflection symmetries about the planes shown in Fig. \ref{ladder} as well as inversion symmetry be present.  As a consequence, the spin-rotational symmetry is not fully broken but retains a residual $U(1)$ symmetry corresponding to a conserved $S_z$ in this model.  From our weak-coupling analysis, we have found that the favored odd-parity state has an in-plane $d$-vector.
We expect that in principle, the Goldstone mode associated with the in-plane spin rotation is gapped due to an explicit symmetry-breaking term (originating either from 
spin-orbit coupling interactions that are ignored in our model 
or from an external in-plane Zeeman field applied in the laboratory). Nevertheless, having incorporated the dominant energy scales into our effective Hamiltonian, it is likely that soft Goldstone modes would be retained to an excellent approximation.

In addition to possessing soft collective excitations, the order parameter considered here can in principle host half-quantum vortices.  Along a closed path that encloses such a defect,  the order parameter
\begin{equation}
\Psi = e^{i \varphi} \Big[ d_x \big( -\vert \uparrow \uparrow \rangle + \vert \downarrow \downarrow \big) + i d_y \big( \vert \uparrow \uparrow \rangle + \vert \downarrow \downarrow \rangle \big)   \Big]
\end{equation}
remains singe-valued when $\varphi \rightarrow \varphi + \pi, \vec d \rightarrow - \vec d$ upon enclosing the defect.  However, such excitations are not favored over ordinary vortices (where $\varphi$ winds by $2 \pi$ without any change in the vector components of $\vec d$) in bulk systems since the spin current is unscreened, leading to a logarithmically divergent energy cost in two dimensions\cite{chung2007}.  These defects, however might exist in mesoscopic samples as is also likely the case in Sr$_2$RuO$_4$\cite{budakian2011}.


With an in-plane $d$-vector, there would be no change in the NMR Knight shift below the superconducting transition, for a field applied along the crystalline b-axis, which is the least resistive transport axis.   
Thus, NMR measurements would be the most direct test of our theory.

Finally, we mention here the role of strong electron interactions.  We have taken on a weak-coupling approach to this system.  However, there are several indications that strong interactions are present in the normal state, including the presence of charge ordering and Luttinger liquid behavior.  Our approach is justified by the fact that at lower temperatures, such Luttinger liquid behavior crosses over into Fermi liquid behavior in this system.  Nonetheless, it will be interesting to study the superconducting instabilities of this system from the vantage point of stronger coupling.  We are currently attempting to do so using density matrix renormalization group calculations on multi-leg ladders, and will report our results in a forthcoming publication.


\begin{acknowledgments}
We are grateful to D. Agterberg and N. Lera for insightful discussions.
This work was supported in part by the Office of Basic Energy Sciences, Materials Sciences and Engineering Division of the US Department of Energy under Contract No. AC02-76SF00515 (WC, SR) (SR).
RHM was supported in part by an Australian Research Council Discovery Project. RT was supported by DFG-SFB 1170, DFG-SPP 1458, and ERC-StG-TOPOLECTRICS.

\end{acknowledgments}

\bibliography{triplet}

\begin{thebibliography}{39}%
\makeatletter
\providecommand \@ifxundefined [1]{%
 \@ifx{#1\undefined}
}%
\providecommand \@ifnum [1]{%
 \ifnum #1\expandafter \@firstoftwo
 \else \expandafter \@secondoftwo
 \fi
}%
\providecommand \@ifx [1]{%
 \ifx #1\expandafter \@firstoftwo
 \else \expandafter \@secondoftwo
 \fi
}%
\providecommand \natexlab [1]{#1}%
\providecommand \enquote  [1]{``#1''}%
\providecommand \bibnamefont  [1]{#1}%
\providecommand \bibfnamefont [1]{#1}%
\providecommand \citenamefont [1]{#1}%
\providecommand \href@noop [0]{\@secondoftwo}%
\providecommand \href [0]{\begingroup \@sanitize@url \@href}%
\providecommand \@href[1]{\@@startlink{#1}\@@href}%
\providecommand \@@href[1]{\endgroup#1\@@endlink}%
\providecommand \@sanitize@url [0]{\catcode `\\12\catcode `\$12\catcode
  `\&12\catcode `\#12\catcode `\^12\catcode `\_12\catcode `\%12\relax}%
\providecommand \@@startlink[1]{}%
\providecommand \@@endlink[0]{}%
\providecommand \url  [0]{\begingroup\@sanitize@url \@url }%
\providecommand \@url [1]{\endgroup\@href {#1}{\urlprefix }}%
\providecommand \urlprefix  [0]{URL }%
\providecommand \Eprint [0]{\href }%
\providecommand \doibase [0]{http://dx.doi.org/}%
\providecommand \selectlanguage [0]{\@gobble}%
\providecommand \bibinfo  [0]{\@secondoftwo}%
\providecommand \bibfield  [0]{\@secondoftwo}%
\providecommand \translation [1]{[#1]}%
\providecommand \BibitemOpen [0]{}%
\providecommand \bibitemStop [0]{}%
\providecommand \bibitemNoStop [0]{.\EOS\space}%
\providecommand \EOS [0]{\spacefactor3000\relax}%
\providecommand \BibitemShut  [1]{\csname bibitem#1\endcsname}%
\let\auto@bib@innerbib\@empty
\bibitem [{\citenamefont {Bao}\ \emph {et~al.}(2015)\citenamefont {Bao},
  \citenamefont {Liu}, \citenamefont {Ma}, \citenamefont {Meng}, \citenamefont
  {Tang}, \citenamefont {Sun}, \citenamefont {Zhai}, \citenamefont {Jiang},
  \citenamefont {Bai}, \citenamefont {Feng} \emph {et~al.}}]{Bao2015}%
  \BibitemOpen
  \bibfield  {author} {\bibinfo {author} {\bibfnamefont {J.-K.}\ \bibnamefont
  {Bao}}, \bibinfo {author} {\bibfnamefont {J.-Y.}\ \bibnamefont {Liu}},
  \bibinfo {author} {\bibfnamefont {C.-W.}\ \bibnamefont {Ma}}, \bibinfo
  {author} {\bibfnamefont {Z.-H.}\ \bibnamefont {Meng}}, \bibinfo {author}
  {\bibfnamefont {Z.-T.}\ \bibnamefont {Tang}}, \bibinfo {author}
  {\bibfnamefont {Y.-L.}\ \bibnamefont {Sun}}, \bibinfo {author} {\bibfnamefont
  {H.-F.}\ \bibnamefont {Zhai}}, \bibinfo {author} {\bibfnamefont
  {H.}~\bibnamefont {Jiang}}, \bibinfo {author} {\bibfnamefont
  {H.}~\bibnamefont {Bai}}, \bibinfo {author} {\bibfnamefont {C.-M.}\
  \bibnamefont {Feng}},  \emph {et~al.},\ }\href@noop {} {\bibfield  {journal}
  {\bibinfo  {journal} {Physical Review X}\ }\textbf {\bibinfo {volume} {5}},\
  \bibinfo {pages} {011013} (\bibinfo {year} {2015})}\BibitemShut {NoStop}%
\bibitem [{\citenamefont {Zhi}\ \emph {et~al.}(2015)\citenamefont {Zhi},
  \citenamefont {Imai}, \citenamefont {Ning}, \citenamefont {Bao},\ and\
  \citenamefont {Cao}}]{Zhi2015}%
  \BibitemOpen
  \bibfield  {author} {\bibinfo {author} {\bibfnamefont {H.~Z.}\ \bibnamefont
  {Zhi}}, \bibinfo {author} {\bibfnamefont {T.}~\bibnamefont {Imai}}, \bibinfo
  {author} {\bibfnamefont {F.~L.}\ \bibnamefont {Ning}}, \bibinfo {author}
  {\bibfnamefont {J.-K.}\ \bibnamefont {Bao}}, \ and\ \bibinfo {author}
  {\bibfnamefont {G.-H.}\ \bibnamefont {Cao}},\ }\href {\doibase
  10.1103/PhysRevLett.114.147004} {\bibfield  {journal} {\bibinfo  {journal}
  {Phys. Rev. Lett.}\ }\textbf {\bibinfo {volume} {114}},\ \bibinfo {pages}
  {147004} (\bibinfo {year} {2015})}\BibitemShut {NoStop}%
\bibitem [{\citenamefont {Lebed}\ \emph {et~al.}(2000)\citenamefont {Lebed},
  \citenamefont {Machida},\ and\ \citenamefont {Ozaki}}]{lebed2000}%
  \BibitemOpen
  \bibfield  {author} {\bibinfo {author} {\bibfnamefont {A.~G.}\ \bibnamefont
  {Lebed}}, \bibinfo {author} {\bibfnamefont {K.}~\bibnamefont {Machida}}, \
  and\ \bibinfo {author} {\bibfnamefont {M.}~\bibnamefont {Ozaki}},\ }\href
  {\doibase 10.1103/PhysRevB.62.R795} {\bibfield  {journal} {\bibinfo
  {journal} {Phys. Rev. B}\ }\textbf {\bibinfo {volume} {62}},\ \bibinfo
  {pages} {R795} (\bibinfo {year} {2000})}\BibitemShut {NoStop}%
\bibitem [{\citenamefont {Zhang}\ and\ \citenamefont
  {S{\'a}~De~Melo}(2007)}]{Zhang2007}%
  \BibitemOpen
  \bibfield  {author} {\bibinfo {author} {\bibfnamefont {W.}~\bibnamefont
  {Zhang}}\ and\ \bibinfo {author} {\bibfnamefont {C.~A.}\ \bibnamefont
  {S{\'a}~De~Melo}},\ }\href@noop {} {\bibfield  {journal} {\bibinfo  {journal}
  {Advances in Physics}\ }\textbf {\bibinfo {volume} {56}},\ \bibinfo {pages}
  {545} (\bibinfo {year} {2007})}\BibitemShut {NoStop}%
\bibitem [{\citenamefont {Maeno}\ \emph {et~al.}(2012)\citenamefont {Maeno},
  \citenamefont {Kittaka}, \citenamefont {Nomura}, \citenamefont {Yonezawa},\
  and\ \citenamefont {Ishida}}]{Maeno2012}%
  \BibitemOpen
  \bibfield  {author} {\bibinfo {author} {\bibfnamefont {Y.}~\bibnamefont
  {Maeno}}, \bibinfo {author} {\bibfnamefont {S.}~\bibnamefont {Kittaka}},
  \bibinfo {author} {\bibfnamefont {T.}~\bibnamefont {Nomura}}, \bibinfo
  {author} {\bibfnamefont {S.}~\bibnamefont {Yonezawa}}, \ and\ \bibinfo
  {author} {\bibfnamefont {K.}~\bibnamefont {Ishida}},\ }\href {\doibase
  10.1143/JPSJ.81.011009} {\bibfield  {journal} {\bibinfo  {journal} {Journal
  of the Physical Society of Japan}\ }\textbf {\bibinfo {volume} {81}},\
  \bibinfo {pages} {011009} (\bibinfo {year} {2012})}\BibitemShut {NoStop}%
\bibitem [{\citenamefont {Strand}\ \emph {et~al.}(2010)\citenamefont {Strand},
  \citenamefont {Bahr}, \citenamefont {Van~Harlingen}, \citenamefont {Davis},
  \citenamefont {Gannon},\ and\ \citenamefont {Halperin}}]{strand2010}%
  \BibitemOpen
  \bibfield  {author} {\bibinfo {author} {\bibfnamefont {J.}~\bibnamefont
  {Strand}}, \bibinfo {author} {\bibfnamefont {D.}~\bibnamefont {Bahr}},
  \bibinfo {author} {\bibfnamefont {D.}~\bibnamefont {Van~Harlingen}}, \bibinfo
  {author} {\bibfnamefont {J.}~\bibnamefont {Davis}}, \bibinfo {author}
  {\bibfnamefont {W.}~\bibnamefont {Gannon}}, \ and\ \bibinfo {author}
  {\bibfnamefont {W.}~\bibnamefont {Halperin}},\ }\href@noop {} {\bibfield
  {journal} {\bibinfo  {journal} {Science}\ }\textbf {\bibinfo {volume}
  {328}},\ \bibinfo {pages} {1368} (\bibinfo {year} {2010})}\BibitemShut
  {NoStop}%
\bibitem [{\citenamefont {Leggett}(1975)}]{Legget75}%
  \BibitemOpen
  \bibfield  {author} {\bibinfo {author} {\bibfnamefont {A.~J.}\ \bibnamefont
  {Leggett}},\ }\href {\doibase 10.1103/RevModPhys.47.331} {\bibfield
  {journal} {\bibinfo  {journal} {Rev. Mod. Phys.}\ }\textbf {\bibinfo {volume}
  {47}},\ \bibinfo {pages} {331} (\bibinfo {year} {1975})}\BibitemShut
  {NoStop}%
\bibitem [{\citenamefont {Sigrist}\ and\ \citenamefont
  {Ueda}(1991)}]{Sigrist91}%
  \BibitemOpen
  \bibfield  {author} {\bibinfo {author} {\bibfnamefont {M.}~\bibnamefont
  {Sigrist}}\ and\ \bibinfo {author} {\bibfnamefont {K.}~\bibnamefont {Ueda}},\
  }\href {\doibase 10.1103/RevModPhys.63.239} {\bibfield  {journal} {\bibinfo
  {journal} {Rev. Mod. Phys.}\ }\textbf {\bibinfo {volume} {63}},\ \bibinfo
  {pages} {239} (\bibinfo {year} {1991})}\BibitemShut {NoStop}%
\bibitem [{\citenamefont {Volovik}\ and\ \citenamefont
  {Gorkov}(1993)}]{volovik1993}%
  \BibitemOpen
  \bibfield  {author} {\bibinfo {author} {\bibfnamefont {G.}~\bibnamefont
  {Volovik}}\ and\ \bibinfo {author} {\bibfnamefont {L.}~\bibnamefont
  {Gorkov}},\ }in\ \href@noop {} {\emph {\bibinfo {booktitle} {Ten Years of
  Superconductivity: 1980--1990}}}\ (\bibinfo  {publisher} {Springer},\
  \bibinfo {year} {1993})\ pp.\ \bibinfo {pages} {144--155}\BibitemShut
  {NoStop}%
\bibitem [{\citenamefont {Mineev}\ and\ \citenamefont
  {Samokhin}(1999)}]{Mineev99}%
  \BibitemOpen
  \bibfield  {author} {\bibinfo {author} {\bibfnamefont {V.}~\bibnamefont
  {Mineev}}\ and\ \bibinfo {author} {\bibfnamefont {K.}~\bibnamefont
  {Samokhin}},\ }\href {https://books.google.com/books?id=2BXYWT8m068C} {\emph
  {\bibinfo {title} {Introduction to Unconventional Superconductivity}}}\
  (\bibinfo  {publisher} {Taylor \& Francis},\ \bibinfo {year}
  {1999})\BibitemShut {NoStop}%
\bibitem [{\citenamefont {Scaffidi}\ \emph {et~al.}(2014)\citenamefont
  {Scaffidi}, \citenamefont {Romers},\ and\ \citenamefont
  {Simon}}]{scaffidi2014}%
  \BibitemOpen
  \bibfield  {author} {\bibinfo {author} {\bibfnamefont {T.}~\bibnamefont
  {Scaffidi}}, \bibinfo {author} {\bibfnamefont {J.~C.}\ \bibnamefont
  {Romers}}, \ and\ \bibinfo {author} {\bibfnamefont {S.~H.}\ \bibnamefont
  {Simon}},\ }\href {\doibase 10.1103/PhysRevB.89.220510} {\bibfield  {journal}
  {\bibinfo  {journal} {Phys. Rev. B}\ }\textbf {\bibinfo {volume} {89}},\
  \bibinfo {pages} {220510} (\bibinfo {year} {2014})}\BibitemShut {NoStop}%
\bibitem [{\citenamefont {Wang}\ \emph {et~al.}(2013)\citenamefont {Wang},
  \citenamefont {Platt}, \citenamefont {Yang}, \citenamefont {Honerkamp},
  \citenamefont {Zhang}, \citenamefont {Hanke}, \citenamefont {Rice},\ and\
  \citenamefont {Thomale}}]{wang2013}%
  \BibitemOpen
  \bibfield  {author} {\bibinfo {author} {\bibfnamefont {Q.~H.}\ \bibnamefont
  {Wang}}, \bibinfo {author} {\bibfnamefont {C.}~\bibnamefont {Platt}},
  \bibinfo {author} {\bibfnamefont {Y.}~\bibnamefont {Yang}}, \bibinfo {author}
  {\bibfnamefont {C.}~\bibnamefont {Honerkamp}}, \bibinfo {author}
  {\bibfnamefont {F.~C.}\ \bibnamefont {Zhang}}, \bibinfo {author}
  {\bibfnamefont {W.}~\bibnamefont {Hanke}}, \bibinfo {author} {\bibfnamefont
  {T.~M.}\ \bibnamefont {Rice}}, \ and\ \bibinfo {author} {\bibfnamefont
  {R.}~\bibnamefont {Thomale}},\ }\href@noop {} {\bibfield  {journal} {\bibinfo
   {journal} {EPL (Europhysics Letters)}\ }\textbf {\bibinfo {volume} {104}},\
  \bibinfo {pages} {17013} (\bibinfo {year} {2013})}\BibitemShut {NoStop}%
\bibitem [{\citenamefont {Yanase}\ \emph {et~al.}(2014)\citenamefont {Yanase},
  \citenamefont {Takamatsu},\ and\ \citenamefont {Udagawa}}]{yanase2014}%
  \BibitemOpen
  \bibfield  {author} {\bibinfo {author} {\bibfnamefont {Y.}~\bibnamefont
  {Yanase}}, \bibinfo {author} {\bibfnamefont {S.}~\bibnamefont {Takamatsu}}, \
  and\ \bibinfo {author} {\bibfnamefont {M.}~\bibnamefont {Udagawa}},\
  }\href@noop {} {\bibfield  {journal} {\bibinfo  {journal} {Journal of the
  Physical Society of Japan}\ }\textbf {\bibinfo {volume} {83}},\ \bibinfo
  {pages} {061019} (\bibinfo {year} {2014})}\BibitemShut {NoStop}%
\bibitem [{\citenamefont {Mercure}\ \emph {et~al.}(2012)\citenamefont
  {Mercure}, \citenamefont {Bangura}, \citenamefont {Xu}, \citenamefont
  {Wakeham}, \citenamefont {Carrington}, \citenamefont {Walmsley},
  \citenamefont {Greenblatt},\ and\ \citenamefont {Hussey}}]{mercure2012}%
  \BibitemOpen
  \bibfield  {author} {\bibinfo {author} {\bibfnamefont {J.-F.}\ \bibnamefont
  {Mercure}}, \bibinfo {author} {\bibfnamefont {A.~F.}\ \bibnamefont
  {Bangura}}, \bibinfo {author} {\bibfnamefont {X.}~\bibnamefont {Xu}},
  \bibinfo {author} {\bibfnamefont {N.}~\bibnamefont {Wakeham}}, \bibinfo
  {author} {\bibfnamefont {A.}~\bibnamefont {Carrington}}, \bibinfo {author}
  {\bibfnamefont {P.}~\bibnamefont {Walmsley}}, \bibinfo {author}
  {\bibfnamefont {M.}~\bibnamefont {Greenblatt}}, \ and\ \bibinfo {author}
  {\bibfnamefont {N.~E.}\ \bibnamefont {Hussey}},\ }\href@noop {} {\bibfield
  {journal} {\bibinfo  {journal} {Physical Review Letters}\ }\textbf {\bibinfo
  {volume} {108}},\ \bibinfo {pages} {187003} (\bibinfo {year}
  {2012})}\BibitemShut {NoStop}%
\bibitem [{\citenamefont {Dudy}\ \emph {et~al.}(2013)\citenamefont {Dudy},
  \citenamefont {Denlinger}, \citenamefont {Allen}, \citenamefont {Wang},
  \citenamefont {He}, \citenamefont {Hitchcock}, \citenamefont {Sekiyama},\
  and\ \citenamefont {Suga}}]{dudy2013}%
  \BibitemOpen
  \bibfield  {author} {\bibinfo {author} {\bibfnamefont {L.}~\bibnamefont
  {Dudy}}, \bibinfo {author} {\bibfnamefont {J.~D.}\ \bibnamefont {Denlinger}},
  \bibinfo {author} {\bibfnamefont {J.~W.}\ \bibnamefont {Allen}}, \bibinfo
  {author} {\bibfnamefont {F.}~\bibnamefont {Wang}}, \bibinfo {author}
  {\bibfnamefont {J.}~\bibnamefont {He}}, \bibinfo {author} {\bibfnamefont
  {D.}~\bibnamefont {Hitchcock}}, \bibinfo {author} {\bibfnamefont
  {A.}~\bibnamefont {Sekiyama}}, \ and\ \bibinfo {author} {\bibfnamefont
  {S.}~\bibnamefont {Suga}},\ }\href@noop {} {\bibfield  {journal} {\bibinfo
  {journal} {Journal of Physics: Condensed Matter}\ }\textbf {\bibinfo {volume}
  {25}},\ \bibinfo {pages} {014007} (\bibinfo {year} {2013})}\BibitemShut
  {NoStop}%
\bibitem [{\citenamefont {Podlich}\ \emph {et~al.}(2013)\citenamefont
  {Podlich}, \citenamefont {Klinke}, \citenamefont {Nansseu}, \citenamefont
  {Waelsch}, \citenamefont {Bienert}, \citenamefont {He}, \citenamefont {Jin},
  \citenamefont {Mandrus},\ and\ \citenamefont {Matzdorf}}]{podlich}%
  \BibitemOpen
  \bibfield  {author} {\bibinfo {author} {\bibfnamefont {T.}~\bibnamefont
  {Podlich}}, \bibinfo {author} {\bibfnamefont {M.}~\bibnamefont {Klinke}},
  \bibinfo {author} {\bibfnamefont {B.}~\bibnamefont {Nansseu}}, \bibinfo
  {author} {\bibfnamefont {M.}~\bibnamefont {Waelsch}}, \bibinfo {author}
  {\bibfnamefont {R.}~\bibnamefont {Bienert}}, \bibinfo {author} {\bibfnamefont
  {J.}~\bibnamefont {He}}, \bibinfo {author} {\bibfnamefont {R.}~\bibnamefont
  {Jin}}, \bibinfo {author} {\bibfnamefont {D.}~\bibnamefont {Mandrus}}, \ and\
  \bibinfo {author} {\bibfnamefont {R.}~\bibnamefont {Matzdorf}},\ }\href@noop
  {} {\bibfield  {journal} {\bibinfo  {journal} {Journal of Physics: Condensed
  Matter}\ }\textbf {\bibinfo {volume} {25}},\ \bibinfo {pages} {014008}
  (\bibinfo {year} {2013})}\BibitemShut {NoStop}%
\bibitem [{\citenamefont {Matsuda}\ \emph {et~al.}(1986)\citenamefont
  {Matsuda}, \citenamefont {Sato}, \citenamefont {Onoda},\ and\ \citenamefont
  {Nakao}}]{matsuda1986}%
  \BibitemOpen
  \bibfield  {author} {\bibinfo {author} {\bibfnamefont {Y.}~\bibnamefont
  {Matsuda}}, \bibinfo {author} {\bibfnamefont {M.}~\bibnamefont {Sato}},
  \bibinfo {author} {\bibfnamefont {M.}~\bibnamefont {Onoda}}, \ and\ \bibinfo
  {author} {\bibfnamefont {K.}~\bibnamefont {Nakao}},\ }\href@noop {}
  {\bibfield  {journal} {\bibinfo  {journal} {Journal of Physics C: Solid State
  Physics}\ }\textbf {\bibinfo {volume} {19}},\ \bibinfo {pages} {6039}
  (\bibinfo {year} {1986})}\BibitemShut {NoStop}%
\bibitem [{\citenamefont {Merino}\ and\ \citenamefont
  {McKenzie}(2012)}]{mckenzie2012prb}%
  \BibitemOpen
  \bibfield  {author} {\bibinfo {author} {\bibfnamefont {J.}~\bibnamefont
  {Merino}}\ and\ \bibinfo {author} {\bibfnamefont {R.~H.}\ \bibnamefont
  {McKenzie}},\ }\href@noop {} {\bibfield  {journal} {\bibinfo  {journal}
  {Physical Review B}\ }\textbf {\bibinfo {volume} {85}},\ \bibinfo {pages}
  {235128} (\bibinfo {year} {2012})}\BibitemShut {NoStop}%
\bibitem [{\citenamefont {Cho}\ \emph {et~al.}(2015)\citenamefont {Cho},
  \citenamefont {Platt}, \citenamefont {McKenzie},\ and\ \citenamefont
  {Raghu}}]{Cho2015}%
  \BibitemOpen
  \bibfield  {author} {\bibinfo {author} {\bibfnamefont {W.}~\bibnamefont
  {Cho}}, \bibinfo {author} {\bibfnamefont {C.}~\bibnamefont {Platt}}, \bibinfo
  {author} {\bibfnamefont {R.~H.}\ \bibnamefont {McKenzie}}, \ and\ \bibinfo
  {author} {\bibfnamefont {S.}~\bibnamefont {Raghu}},\ }\href {\doibase
  10.1103/PhysRevB.92.134514} {\bibfield  {journal} {\bibinfo  {journal} {Phys.
  Rev. B}\ }\textbf {\bibinfo {volume} {92}},\ \bibinfo {pages} {134514}
  (\bibinfo {year} {2015})}\BibitemShut {NoStop}%
\bibitem [{\citenamefont {Lera}\ and\ \citenamefont
  {Alvarez}(2015)}]{Lera2015}%
  \BibitemOpen
  \bibfield  {author} {\bibinfo {author} {\bibfnamefont {N.}~\bibnamefont
  {Lera}}\ and\ \bibinfo {author} {\bibfnamefont {J.~V.}\ \bibnamefont
  {Alvarez}},\ }\href {\doibase 10.1103/PhysRevB.92.174523} {\bibfield
  {journal} {\bibinfo  {journal} {Phys. Rev. B}\ }\textbf {\bibinfo {volume}
  {92}},\ \bibinfo {pages} {174523} (\bibinfo {year} {2015})}\BibitemShut
  {NoStop}%
\bibitem [{\citenamefont {Chudzinski}\ \emph {et~al.}(2012)\citenamefont
  {Chudzinski}, \citenamefont {Jarlborg},\ and\ \citenamefont
  {Giamarchi}}]{chudzinski2012}%
  \BibitemOpen
  \bibfield  {author} {\bibinfo {author} {\bibfnamefont {P.}~\bibnamefont
  {Chudzinski}}, \bibinfo {author} {\bibfnamefont {T.}~\bibnamefont
  {Jarlborg}}, \ and\ \bibinfo {author} {\bibfnamefont {T.}~\bibnamefont
  {Giamarchi}},\ }\href@noop {} {\bibfield  {journal} {\bibinfo  {journal}
  {Physical Review B}\ }\textbf {\bibinfo {volume} {86}},\ \bibinfo {pages}
  {075147} (\bibinfo {year} {2012})}\BibitemShut {NoStop}%
\bibitem [{\citenamefont {Nuss}\ and\ \citenamefont
  {Aichhorn}(2014)}]{nuss2014}%
  \BibitemOpen
  \bibfield  {author} {\bibinfo {author} {\bibfnamefont {M.}~\bibnamefont
  {Nuss}}\ and\ \bibinfo {author} {\bibfnamefont {M.}~\bibnamefont
  {Aichhorn}},\ }\href@noop {} {\bibfield  {journal} {\bibinfo  {journal}
  {Physical Review B}\ }\textbf {\bibinfo {volume} {89}},\ \bibinfo {pages}
  {045125} (\bibinfo {year} {2014})}\BibitemShut {NoStop}%
\bibitem [{\citenamefont {Fischer}\ \emph {et~al.}(2011)\citenamefont
  {Fischer}, \citenamefont {Loder},\ and\ \citenamefont
  {Sigrist}}]{Fischer2011}%
  \BibitemOpen
  \bibfield  {author} {\bibinfo {author} {\bibfnamefont {M.~H.}\ \bibnamefont
  {Fischer}}, \bibinfo {author} {\bibfnamefont {F.}~\bibnamefont {Loder}}, \
  and\ \bibinfo {author} {\bibfnamefont {M.}~\bibnamefont {Sigrist}},\
  }\href@noop {} {\bibfield  {journal} {\bibinfo  {journal} {Physical Review
  B}\ }\textbf {\bibinfo {volume} {84}},\ \bibinfo {pages} {184533} (\bibinfo
  {year} {2011})}\BibitemShut {NoStop}%
\bibitem [{\citenamefont {Maruyama}\ \emph {et~al.}(2012)\citenamefont
  {Maruyama}, \citenamefont {Sigrist},\ and\ \citenamefont
  {Yanase}}]{maruyama2012}%
  \BibitemOpen
  \bibfield  {author} {\bibinfo {author} {\bibfnamefont {D.}~\bibnamefont
  {Maruyama}}, \bibinfo {author} {\bibfnamefont {M.}~\bibnamefont {Sigrist}}, \
  and\ \bibinfo {author} {\bibfnamefont {Y.}~\bibnamefont {Yanase}},\
  }\href@noop {} {\bibfield  {journal} {\bibinfo  {journal} {Journal of the
  Physical Society of Japan}\ }\textbf {\bibinfo {volume} {81}},\ \bibinfo
  {pages} {034702} (\bibinfo {year} {2012})}\BibitemShut {NoStop}%
\bibitem [{\citenamefont {Kane}\ and\ \citenamefont {Mele}(2005)}]{Kane2005}%
  \BibitemOpen
  \bibfield  {author} {\bibinfo {author} {\bibfnamefont {C.~L.}\ \bibnamefont
  {Kane}}\ and\ \bibinfo {author} {\bibfnamefont {E.~J.}\ \bibnamefont
  {Mele}},\ }\href@noop {} {\bibfield  {journal} {\bibinfo  {journal} {Physical
  Review Letters}\ }\textbf {\bibinfo {volume} {95}},\ \bibinfo {pages}
  {226801} (\bibinfo {year} {2005})}\BibitemShut {NoStop}%
\bibitem [{\citenamefont {Min}\ \emph {et~al.}(2006)\citenamefont {Min},
  \citenamefont {Hill}, \citenamefont {Sinitsyn}, \citenamefont {Sahu},
  \citenamefont {Kleinman},\ and\ \citenamefont {Macdonald}}]{Min2006}%
  \BibitemOpen
  \bibfield  {author} {\bibinfo {author} {\bibfnamefont {H.}~\bibnamefont
  {Min}}, \bibinfo {author} {\bibfnamefont {J.~E.}\ \bibnamefont {Hill}},
  \bibinfo {author} {\bibfnamefont {N.~A.}\ \bibnamefont {Sinitsyn}}, \bibinfo
  {author} {\bibfnamefont {B.~R.}\ \bibnamefont {Sahu}}, \bibinfo {author}
  {\bibfnamefont {L.}~\bibnamefont {Kleinman}}, \ and\ \bibinfo {author}
  {\bibfnamefont {A.~H.}\ \bibnamefont {Macdonald}},\ }\href@noop {} {\bibfield
   {journal} {\bibinfo  {journal} {Physical Review B}\ }\textbf {\bibinfo
  {volume} {74}},\ \bibinfo {pages} {165310} (\bibinfo {year}
  {2006})}\BibitemShut {NoStop}%
\bibitem [{\citenamefont {Xiao}\ \emph {et~al.}(2012)\citenamefont {Xiao},
  \citenamefont {Liu}, \citenamefont {Feng}, \citenamefont {Xu},\ and\
  \citenamefont {Yao}}]{Xiao2012}%
  \BibitemOpen
  \bibfield  {author} {\bibinfo {author} {\bibfnamefont {D.}~\bibnamefont
  {Xiao}}, \bibinfo {author} {\bibfnamefont {G.-B.}\ \bibnamefont {Liu}},
  \bibinfo {author} {\bibfnamefont {W.}~\bibnamefont {Feng}}, \bibinfo {author}
  {\bibfnamefont {X.}~\bibnamefont {Xu}}, \ and\ \bibinfo {author}
  {\bibfnamefont {W.}~\bibnamefont {Yao}},\ }\href@noop {} {\bibfield
  {journal} {\bibinfo  {journal} {Phys. Rev. Lett.}\ }\textbf {\bibinfo
  {volume} {108}},\ \bibinfo {pages} {196802} (\bibinfo {year}
  {2012})}\BibitemShut {NoStop}%
\bibitem [{\citenamefont {Youn}\ \emph {et~al.}(2012)\citenamefont {Youn},
  \citenamefont {Fischer}, \citenamefont {Rhim}, \citenamefont {Sigrist},\ and\
  \citenamefont {Agterberg}}]{Youn2012}%
  \BibitemOpen
  \bibfield  {author} {\bibinfo {author} {\bibfnamefont {S.~J.}\ \bibnamefont
  {Youn}}, \bibinfo {author} {\bibfnamefont {M.~H.}\ \bibnamefont {Fischer}},
  \bibinfo {author} {\bibfnamefont {S.~H.}\ \bibnamefont {Rhim}}, \bibinfo
  {author} {\bibfnamefont {M.}~\bibnamefont {Sigrist}}, \ and\ \bibinfo
  {author} {\bibfnamefont {D.~F.}\ \bibnamefont {Agterberg}},\ }\href@noop {}
  {\bibfield  {journal} {\bibinfo  {journal} {Physical Review B}\ }\textbf
  {\bibinfo {volume} {85}},\ \bibinfo {pages} {220505} (\bibinfo {year}
  {2012})}\BibitemShut {NoStop}%
\bibitem [{\citenamefont {Fischer}\ \emph {et~al.}(2014)\citenamefont
  {Fischer}, \citenamefont {Neupert}, \citenamefont {Platt}, \citenamefont
  {Schnyder}, \citenamefont {Hanke}, \citenamefont {Goryo}, \citenamefont
  {Thomale},\ and\ \citenamefont {Sigrist}}]{PhysRevB.89.020509}%
  \BibitemOpen
  \bibfield  {author} {\bibinfo {author} {\bibfnamefont {M.~H.}\ \bibnamefont
  {Fischer}}, \bibinfo {author} {\bibfnamefont {T.}~\bibnamefont {Neupert}},
  \bibinfo {author} {\bibfnamefont {C.}~\bibnamefont {Platt}}, \bibinfo
  {author} {\bibfnamefont {A.~P.}\ \bibnamefont {Schnyder}}, \bibinfo {author}
  {\bibfnamefont {W.}~\bibnamefont {Hanke}}, \bibinfo {author} {\bibfnamefont
  {J.}~\bibnamefont {Goryo}}, \bibinfo {author} {\bibfnamefont
  {R.}~\bibnamefont {Thomale}}, \ and\ \bibinfo {author} {\bibfnamefont
  {M.}~\bibnamefont {Sigrist}},\ }\href {\doibase 10.1103/PhysRevB.89.020509}
  {\bibfield  {journal} {\bibinfo  {journal} {Phys. Rev. B}\ }\textbf {\bibinfo
  {volume} {89}},\ \bibinfo {pages} {020509} (\bibinfo {year}
  {2014})}\BibitemShut {NoStop}%
\bibitem [{\citenamefont {Annett}(1990)}]{annett1990}%
  \BibitemOpen
  \bibfield  {author} {\bibinfo {author} {\bibfnamefont {J.~F.}\ \bibnamefont
  {Annett}},\ }\href@noop {} {\bibfield  {journal} {\bibinfo  {journal}
  {Advances in Physics}\ }\textbf {\bibinfo {volume} {39}},\ \bibinfo {pages}
  {83} (\bibinfo {year} {1990})}\BibitemShut {NoStop}%
\bibitem [{\citenamefont {Raghu}\ \emph
  {et~al.}(2010{\natexlab{a}})\citenamefont {Raghu}, \citenamefont {Kivelson},\
  and\ \citenamefont {Scalapino}}]{raghu2010weakcoupling}%
  \BibitemOpen
  \bibfield  {author} {\bibinfo {author} {\bibfnamefont {S.}~\bibnamefont
  {Raghu}}, \bibinfo {author} {\bibfnamefont {S.}~\bibnamefont {Kivelson}}, \
  and\ \bibinfo {author} {\bibfnamefont {D.}~\bibnamefont {Scalapino}},\
  }\href@noop {} {\bibfield  {journal} {\bibinfo  {journal} {Physical Review
  B}\ }\textbf {\bibinfo {volume} {81}},\ \bibinfo {pages} {224505} (\bibinfo
  {year} {2010}{\natexlab{a}})}\BibitemShut {NoStop}%
\bibitem [{\citenamefont {Raghu}\ \emph
  {et~al.}(2010{\natexlab{b}})\citenamefont {Raghu}, \citenamefont
  {Kapitulnik},\ and\ \citenamefont {Kivelson}}]{Raghu2010}%
  \BibitemOpen
  \bibfield  {author} {\bibinfo {author} {\bibfnamefont {S.}~\bibnamefont
  {Raghu}}, \bibinfo {author} {\bibfnamefont {A.}~\bibnamefont {Kapitulnik}}, \
  and\ \bibinfo {author} {\bibfnamefont {S.~A.}\ \bibnamefont {Kivelson}},\
  }\href {\doibase 10.1103/PhysRevLett.105.136401} {\bibfield  {journal}
  {\bibinfo  {journal} {Phys. Rev. Lett.}\ }\textbf {\bibinfo {volume} {105}},\
  \bibinfo {pages} {136401} (\bibinfo {year} {2010}{\natexlab{b}})}\BibitemShut
  {NoStop}%
\bibitem [{\citenamefont {Kiesel}\ and\ \citenamefont
  {Thomale}(2012)}]{PhysRevB.86.121105}%
  \BibitemOpen
  \bibfield  {author} {\bibinfo {author} {\bibfnamefont {M.~L.}\ \bibnamefont
  {Kiesel}}\ and\ \bibinfo {author} {\bibfnamefont {R.}~\bibnamefont
  {Thomale}},\ }\href {\doibase 10.1103/PhysRevB.86.121105} {\bibfield
  {journal} {\bibinfo  {journal} {Phys. Rev. B}\ }\textbf {\bibinfo {volume}
  {86}},\ \bibinfo {pages} {121105} (\bibinfo {year} {2012})}\BibitemShut
  {NoStop}%
\bibitem [{\citenamefont {Cho}\ \emph {et~al.}(2013)\citenamefont {Cho},
  \citenamefont {Thomale}, \citenamefont {Raghu},\ and\ \citenamefont
  {Kivelson}}]{cho2013}%
  \BibitemOpen
  \bibfield  {author} {\bibinfo {author} {\bibfnamefont {W.}~\bibnamefont
  {Cho}}, \bibinfo {author} {\bibfnamefont {R.}~\bibnamefont {Thomale}},
  \bibinfo {author} {\bibfnamefont {S.}~\bibnamefont {Raghu}}, \ and\ \bibinfo
  {author} {\bibfnamefont {S.~A.}\ \bibnamefont {Kivelson}},\ }\href@noop {}
  {\bibfield  {journal} {\bibinfo  {journal} {Physical Review B}\ }\textbf
  {\bibinfo {volume} {88}},\ \bibinfo {pages} {064505} (\bibinfo {year}
  {2013})}\BibitemShut {NoStop}%
\bibitem [{\citenamefont {Kohn}\ and\ \citenamefont
  {Luttinger}(1965)}]{kohn-65prl524}%
  \BibitemOpen
  \bibfield  {author} {\bibinfo {author} {\bibfnamefont {W.}~\bibnamefont
  {Kohn}}\ and\ \bibinfo {author} {\bibfnamefont {J.~M.}\ \bibnamefont
  {Luttinger}},\ }\href@noop {} {\bibfield  {journal} {\bibinfo  {journal}
  {Physical Review Letters}\ }\textbf {\bibinfo {volume} {15}},\ \bibinfo
  {pages} {524} (\bibinfo {year} {1965})}\BibitemShut {NoStop}%
\bibitem [{\citenamefont {Polchinski}(1992)}]{Polchinski99}%
  \BibitemOpen
  \bibfield  {author} {\bibinfo {author} {\bibfnamefont {J.}~\bibnamefont
  {Polchinski}},\ }in\ \href@noop {} {\emph {\bibinfo {booktitle} {{Theoretical
  Advanced Study Institute (TASI 92): From Black Holes and Strings to Particles
  Boulder, Colorado, June 3-28, 1992}}}}\ (\bibinfo {year} {1992})\ \Eprint
  {http://arxiv.org/abs/hep-th/9210046} {arXiv:hep-th/9210046 [hep-th]}
  \BibitemShut {NoStop}%
\bibitem [{\citenamefont {Shankar}(1994)}]{Shankar94}%
  \BibitemOpen
  \bibfield  {author} {\bibinfo {author} {\bibfnamefont {R.}~\bibnamefont
  {Shankar}},\ }\href@noop {} {\bibfield  {journal} {\bibinfo  {journal} {Rev.
  Mod. Phys.}\ }\textbf {\bibinfo {volume} {66}},\ \bibinfo {pages} {129}
  (\bibinfo {year} {1994})}\BibitemShut {NoStop}%
\bibitem [{\citenamefont {Chung}\ \emph {et~al.}(2007)\citenamefont {Chung},
  \citenamefont {Bluhm},\ and\ \citenamefont {Kim}}]{chung2007}%
  \BibitemOpen
  \bibfield  {author} {\bibinfo {author} {\bibfnamefont {S.~B.}\ \bibnamefont
  {Chung}}, \bibinfo {author} {\bibfnamefont {H.}~\bibnamefont {Bluhm}}, \ and\
  \bibinfo {author} {\bibfnamefont {E.-A.}\ \bibnamefont {Kim}},\ }\href
  {\doibase 10.1103/PhysRevLett.99.197002} {\bibfield  {journal} {\bibinfo
  {journal} {Phys. Rev. Lett.}\ }\textbf {\bibinfo {volume} {99}},\ \bibinfo
  {pages} {197002} (\bibinfo {year} {2007})}\BibitemShut {NoStop}%
\bibitem [{\citenamefont {Jang}\ \emph {et~al.}(2011)\citenamefont {Jang},
  \citenamefont {Ferguson}, \citenamefont {Vakaryuk}, \citenamefont {Budakian},
  \citenamefont {Chung}, \citenamefont {Goldbart},\ and\ \citenamefont
  {Maeno}}]{budakian2011}%
  \BibitemOpen
  \bibfield  {author} {\bibinfo {author} {\bibfnamefont {J.}~\bibnamefont
  {Jang}}, \bibinfo {author} {\bibfnamefont {D.~G.}\ \bibnamefont {Ferguson}},
  \bibinfo {author} {\bibfnamefont {V.}~\bibnamefont {Vakaryuk}}, \bibinfo
  {author} {\bibfnamefont {R.}~\bibnamefont {Budakian}}, \bibinfo {author}
  {\bibfnamefont {S.~B.}\ \bibnamefont {Chung}}, \bibinfo {author}
  {\bibfnamefont {P.~M.}\ \bibnamefont {Goldbart}}, \ and\ \bibinfo {author}
  {\bibfnamefont {Y.}~\bibnamefont {Maeno}},\ }\href@noop {} {\bibfield
  {journal} {\bibinfo  {journal} {Science}\ }\textbf {\bibinfo {volume}
  {331}},\ \bibinfo {pages} {186} (\bibinfo {year} {2011})}\BibitemShut
  {NoStop}%
\end{thebibliography}%

\end{document}